\documentclass[aps,prl,twocolumn,showpacs,superscriptaddress]{revtex4-1}
\usepackage{amsmath}
\usepackage{graphicx}
\usepackage{amsfonts}
\usepackage{amssymb}

\begin{document}

\title{The origin of magnetic freezing in the pyrochlore Y$_{2}$Mo$_{2}$O$_{7}$}
\author{Oren~Ofer}
\email{oren@triumf.ca}
\affiliation{TRIUMF, 4004 Wesbrook Mall, Vancouver, BC, V6T 2A3 Canada} %
\affiliation{Department of Physics, Technion, Haifa 32000, Israel}
\author{Amit Keren}
\affiliation{Department of Physics, Technion, Haifa 32000, Israel}
\author{Jason S. Gardner}
\affiliation{NCNR, National Institute of Standards and Technology, Gaithersburg, MD
20899-6102} \affiliation{Indiana University, 2401 Milo B. Sampson Lane,
Bloomington, IN 47408}
\author{Yang Ren}
\affiliation{Advanced Photon Source, Argonne National Laboratory, 9700 South
Cass Ave., Argonne, IL 60439 USA}
\author{W. A. MacFarlane}
\affiliation{Department of Chemistry,  University of British Columbia,
Vancouver, BC, V6T 1Z1 Canada}

\date{\today}

\begin{abstract}
We investigated the nature of the spin glass-like phase transition in the
geometrically frustrated pyrochlore lattices Y$_{2}$Mo$_{2}$O$_{7}$ using
the local probes nuclear and muon magnetic resonances, and the
field-dependent long range probes x-ray and neutron scattering. The long
range probes indicated that Y$_{2}$Mo$_{2}$O$_{7}$ does not undergo any
global symmetry changes, even in a field of $6$~T. In contrast, the local
signal indicates a lattice distortion close to the critical temperature. The
nuclei show at least two inequivalent Y sites, and the muons show sub-linear
line broadening as a function of moment size, over a wide temperature range. The conclusion from all the
measurements is that even in high field, the distortion of Y$_{2}$Mo$_{2}$O$%
_{7}$ takes place within the unit-cell, while its global cubic symmetry is
preserved. Moreover, the muon result clearly indicates the presence of
magneto-elastic coupling.
\end{abstract}

\maketitle

The Heisenberg model on the geometrically frustrated pyrochlore lattice has
a macroscopically degenerate ground state, and the standard
degeneracy-lifting mechanism of thermal or quantum fluctuation does not seem
to remove it. Yet, with only one exception, the family of compounds based on
Tb$_{2}$Ti$_{2}$O$_{7}$ \cite{jason,tbtio,isabel,ueland}, all pyrochlores
freeze (at least partially), that is, one state out of many is selected. In
some cases, like the spin ice, this is due to long-range interactions \cite%
{yavorskii} and single ion anisotropy \cite{spinice,henley}. In others, the
freezing occurs even without anisotropy. In these cases magneto-elastic
coupling might be responsible for the degeneracy lifting; the lattice
distorts to relieve the frustrated interactions. Such a distortion might
lead to a cubic-to-lower-symmetry structural transition \cite{yamashita}.
This kind of frustration-driven distortion has been previously suggested as
the main freezing mechanism for several $A_{2}B_{2}O_{7}$ pyrochlores \cite%
{amit1,eva} and Cr spinels \cite{lee,chung}, and was considered
theoretically \cite{oleg,doron,penc,tim}. However, lattice distortions and
symmetry changes as a function of temperature are a common feature in
solids, even without magnetic interactions. Therefore, it is not yet clear
whether: (I) the magnetic interactions drive the distortion; or (II) the
distortion takes place because of electrostatic interactions, and the
magnetic properties, such as freezing, follow. Clarification of this point
is crucial for the understanding of the spin Hamiltonian, and therefore the
ground state and excitations in pyrochlores. Field-dependence experiments
can provide, in principle, answers to these questions.
\begin{figure}[b]
\includegraphics[width=0.75\columnwidth]{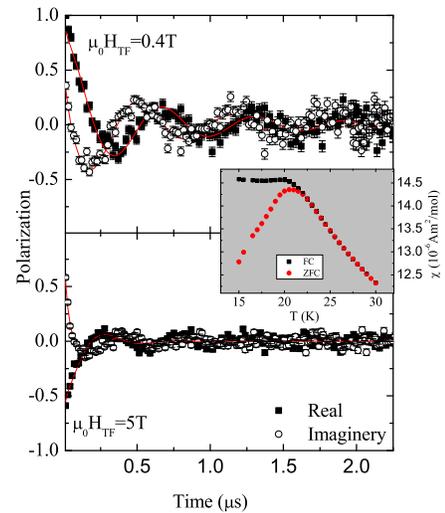}
\caption{(Color online) The raw muon depolarization data taken at $T=24$K,
at two transverse fields. The error bars represent statistical errors. The
inset shows the bulk susceptibility $\protect\chi =M/H$ versus temperature
in zero-field cooled and field cooled conditions taken at a 1kG field.}
\label{fig1:rawMuSR}
\end{figure}

One case is the pyrochlore Y$_{2}$Mo$_{2}$O$_{7}$ (YMO), which crystallize
into a cubic structure with Fd$\overline{3}$m symmetry \cite{reimers}.
Magnetically, it has a Curie-Weiss temperature of $-200$~K and freezes with
spin-glass characteristics \cite{gingras} at $T_{f}=22.5$~K \cite{raju}. In
particular, magnetization measurements indicate a large difference between
zero-field-cooled and field-cooled magnetizations \cite{Miyoshi,
GreedanYMOsusc}. This glassiness is unexpected if the chemical structure is
perfect because it is believed that a spin-glass state emerges when
frustration \textit{and} disorder coexist. A possible solution to this
dichotomy came from more detailed measurements. X-ray absorption fine
structure spectroscopy (XAFS) in zero field show evidence of positional
disorder of the Mo ion \cite{booth}. In contrast, neutron pair distribution
function measurements assigned the distortion to the O1-Y bond \cite{Greedan}%
. Local magnetic probes, which are coupled to the spin system and operate in
a field, such as $^{89}$Y nuclear magnetic resonance (NMR) \cite{amit1} and
muon spin relaxation ($\mu $SR) \cite{eva}, indicate a distortion of the
spin-bearing ion, again the Mo. Despite the controversy on the distorted
ion, all experiments suggest the lattice \textit{is} taking part in the
magnetic freezing. However, the presence of an applied field in the
resonance measurements and its absence in scattering measurements make the
comparison difficult. Moreover, neither experiment can shed light on the
magneto-elastic coupling issue discussed above. The motivation of this work
is to fill in the gap and perform both local resonance and long range
scattering measurements under the same conditions, and to use the field as a
probe of magneto-elastic coupling.

\begin{figure}[tbp]
\includegraphics[width=\columnwidth]{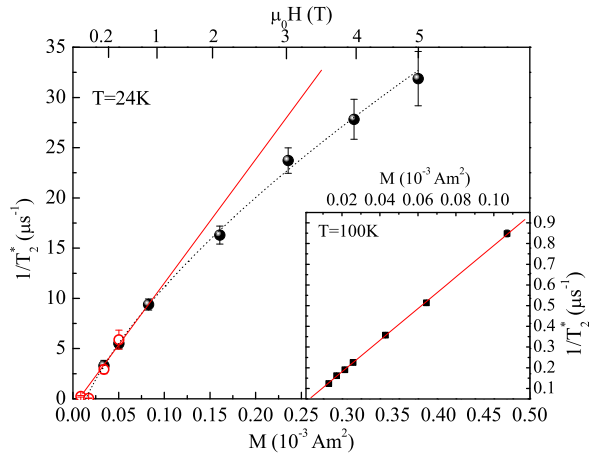}
\caption{(Color online) The TF-$\protect\mu $SR relaxation rate, $%
1/T_{2}^{\ast }$, versus the magnetization, $M$ at $T=24$K. The field $%
\protect\mu _{0}H$ is an implicit parameter. Hollow circles are taken from
Ref.\protect\cite{eva}. The red (solid) line is a linear fit to low $M$
data. The black (dotted) line is a fit to $A\cdot M^{\protect\beta }$ where $%
\protect\beta =0.66(12)$. The inset shows $1/T_{2}^{\ast }$ versus $M$ at $%
T=100$~K.}
\label{fig:t2vsm}
\end{figure}

We performed five different experiments on YMO, which were carried out well
above and close to $T_{f}$. (i) High transverse field (TF) and longitudinal
field (LF) $\mu $SR, which extends previously low-field data \cite%
{eva,dunsiger}. (ii) $^{89}$Y NMR where we extend previous data \cite{amit1}
to the helium range. (iii) Field-dependent high resolution x-ray powder
diffraction. (iv) Field-dependent neutrons diffraction, which extends the
previous zero field (ZF) measurement \cite{reimers,Greedan}. (v) Bulk
magnetization using a superconducting quantum interference device (SQUID).

Polycrystalline samples of YMO were prepared according to Ref.~\cite{raju}.
The inset of Fig.~\ref{fig1:rawMuSR} shows the zero-field-cooled (ZFC) and
field-cooled (FC) magnetization measurements. The ZFC curve shows a
distinctive maximum indicating the spin-glass transition. $T_{f}$ and
Curie-Weiss temperatures extracted from these measurements are in agreement
with previous reports. In addition magnetization vs. applied field up to $5$%
~T at various temperatures was also recorded and will be used below.

\begin{figure}[tb]
\includegraphics[width=\columnwidth]{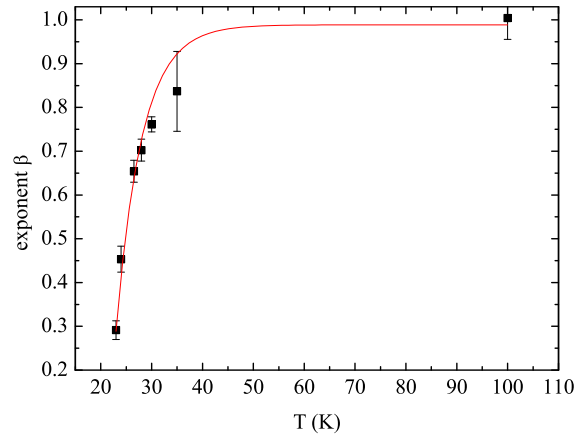}
\caption{(Color online) The exponent $\beta$ versus the Temperature. $\beta$ is extracted from $1/T_2^\ast$ versus the magnetization $M$ fits to $1/T_2^\ast=AM^\beta$. Line is a guide to the eye.}
\label{fig:betaVsT}
\end{figure}

$\mu $SR measurements were performed at TRIUMF, Canada on the HiTime and
Helios spectrometers in the M15 surface muon channel. Figure \ref%
{fig1:rawMuSR} depicts the raw muon polarization taken at a constant
temperature, $T=24$~K, at two transverse fields $\mu _{0}H_{\text{TF}}=0.4$
and $5$~T. The data is shown in a rotating-reference frame \cite{jessRRF} of
$\mu _{0}H_{\text{RRF}}=\mu _{0}H_{\text{TF}}-0.02$~T. The TF relaxation
rate increases with increasing fields. The LF (not shown) is at least an
order of magnitude smaller than the TF relaxation rate at any field and
temperature. This means the muon depolarization is mainly due to static
field inhomogeneities and the contribution of the out-of-plane
depolarization can be neglected. We found that the $\mu $SR TF asymmetry is
best described by
\begin{equation}
P_{\text{TF}}(t)=P_{0}\exp \left(-\sqrt{t/T_{2}^{\ast }}\right)\cos (\omega t+\varphi )
\end{equation}%
where $P_{0}$ is the initial polarization and $\omega =\gamma _{\mu }H_{\text{TF%
}}$. The fits are represented by the solid line in Fig.~\ref{fig1:rawMuSR}.

In Fig.~\ref{fig:t2vsm} we plot the transverse field relaxation rate $%
1/T_{2}^{\ast }$ at a given field versus the magnetization $M$ measured at
the same field, at temperature fixed at $24~$K, slightly above the
spin-glass freezing. We also show the field, $H$, as an explicit parameter
on the upper abscissa. As the field increases the relaxation rate $%
1/T_{2}^{\ast }$ increases. However above $M=10^{-4}$~Am$^2$, $1/T_{2}^{\ast
}$ is no longer a linear function of $M$. This is not the case at higher
temperatures. For comparison, in the inset of Fig.~\ref{fig:t2vsm} we plot $%
1/T_{2}^{\ast }$ vs $M$ taken at $100$~K, again with $H$ as an implicit
parameter. At this temperature the relaxation rate is a linear function of $M
$.

In a system with quenched disorder it is expected that $1/T_{2}^{\ast
}\propto M$ \cite{tbtio,KerenChptr} as is indeed observed at $100$~K.
However, when $T $ approaches $T_{f}$, the data deviates from this linearity
and $1/T_{2}^{\ast }$ grows more slowly than $M$ as depicted in Fig.~\ref%
{fig:t2vsm}. Fits to $1/T_{2}^{\ast }=AM^{\beta }$ are also shown in the
figure, giving $\beta =0.66(12)$ and $\beta=1.00(2)$ for $T=24$ and $100$~K respectively.
It is found that the transition from $\beta=1$ to very low values are spread over a wide temperature range (Fig.~\ref{fig:betaVsT}).

One possibility is a site-dependent spin polarization due to impurities \cite%
{alloul}. However, this can be ruled out by the undetectable amount of
disorder in the sample. A more likely scenario is that the disorder in YMO
is not quenched and at low $T$ it varies with the field. In fact, the
sub-linear behavior suggests that as the field (and $M$) increases the
lattice is more ordered. This is the main finding of this work, which
clearly points to the presence of a magneto-elastic coupling.

\begin{figure}[tbp]
\includegraphics[width=\columnwidth]{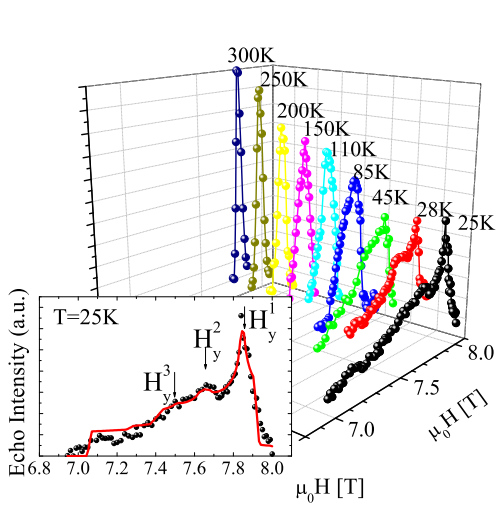}
\caption{(Color online) The NMR spectra at $25\leq T\leq 300$~K. Inset: the
NMR spectrum at $T=25$~K. The line is a fit to a powder spectrum with
three sites. The arrows point to the main peak in the powder spectrum of
each site.}
\label{fig:nmr}
\end{figure}

More evidence for temperature dependent disorder comes from $^{89}$Y NMR
which extends previous measurements \cite{amit1} down to $25$~K. Such low-$T$
measurements were made possible by using a high pressure cell where higher
RF power can be delivered to the sample. At each temperature we obtain the
complete NMR spectrum by sweeping the external field, $H_{ext}$, at a
constant applied RF frequency $f=16.44$~MHz. In each field we used the
spin-echo sequence ($\pi /2-\pi $ pulses) and recorded the echo signal. In
Fig.~\ref{fig:nmr} we present the spectra taken at temperatures between $300$%
~K and $25$~K. The width of the $300$~K spectrum extends over $0.1$~T,
whereas the width of the $25$~K spectrum extends over $1~$T. This broadening
results in low intensities at each applied field upon cooling. Due to this
broad line at low $T$ we gave up on high resolution NMR, as in Ref.~\cite%
{amit1}, and concentrated on the gross features of the spectrum. The most
noticeable feature in the $T=25$~K spectrum is the clear appearance of two
peaks, with a hint of a third one. This suggests that out of the many different Y sites existing at high $T$\cite{amit1}, only few are being picked as $T$ is lowered.

\begin{figure}[tbp]
\includegraphics[width=\columnwidth]{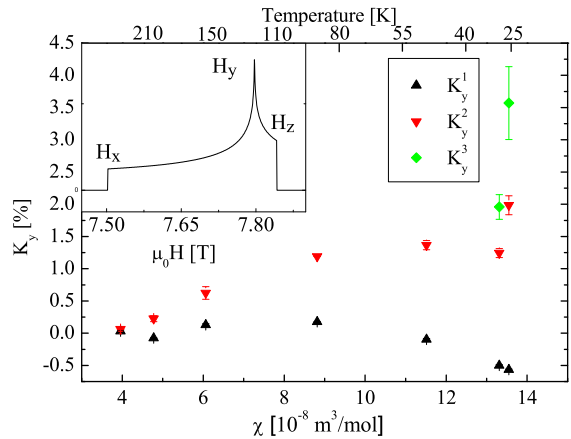}
\caption{(Color online) The NMR shift, $K_{y}$, versus the bulk
susceptiblity $\protect\chi $ and the temperature as an implicit parameter.
In the inset, a theoretical powder spectrum of a nuclear spin $1/2$ is
depicted.}
\label{fig:kvsChi}
\end{figure}

To understand this NMR spectrum we look at the spin $1/2$ $^{89}$Y nuclear
spin Hamiltonian, which can be described as,
\begin{equation}
\mathcal{H}=-\hbar \gamma \mathbf{I}{}\cdot \left( \mathbf{\overline{1}}+%
\mathbf{\overline{K}}\right) \cdot \mathbf{H}_{ext}
\end{equation}%
where $\mathbf{{\overline{K}}}$ is the NMR shift tensor, $\mathbf{I}$ is the
nuclear spin operator, and $\mathbf{H}_{ext}$ is the variable applied field.
In powders, the principal axes of $\mathbf{{\overline{K}}}$ are randomly
oriented relative to $\mathbf{H}_{ext}$. Therefore, the magnetic resonance
spectrum is an average over all possible orientations. A theoretical powder
averaged NMR line \cite{taylor}, for a single site, is depicted in the inset
of Fig.~\ref{fig:kvsChi} where $H_{\alpha }=2\pi f/[\gamma (1+K_{\alpha })]$
and $\alpha =x,y,z$ for the three directions. This theoretical spectrum
demonstrates that a single site, with a single set of $K_{x}$, $K_{y}$, $%
K_{z}$, could give rise to only one peak even under powder averaging. The
existence of two (perhaps three) peaks in the spectrum is a result of a
lattice deformation leading to inequivalent Y sites.

In order to study the temperature dependence of the shift, we use the powder
spectrum convoluted with Lorentzians to fit the NMR spectra. Such a fit is
demonstrated by the solid line in the inset of Fig.~\ref{fig:nmr}. At high
temperatures ($T>250$~K) the shift is very small; at intermediate
temperatures, $50\leq T\leq 200$~K, two different sites were needed to fit
the data (see Fig.~\ref{fig:nmr}); and finally, at low enough temperatures, $%
T<50$~K, three sites were assumed. In Fig.~\ref{fig:kvsChi} we plot the
shift $K_{y}^{i}$, $i$ for each site, versus the bulk susceptibility, which
was extrapolated to $\mu_0H=7.8$~T from the FC magnetization measurements.
Temperature in this figure is an implicit parameter. As $\chi$ increases the
shift for each site also increases. However, the dependence between shift
and susceptibility is not linear. When the disorder is quenched one expects $%
K\propto \chi $. This proportionality is violated close to $T_{f}$,
indicating that the lattice degrees of freedom are active as $T\rightarrow
T_{f}$. It should be pointed out that unlike in $\mu $SR, in NMR it is
impossible to vary the field over a wide range and NMR cannot be used to
address the question of magneto-elastic coupling. In contrast, the presence of multiple sites well above $T_f$ suggest that spin correlations are sufficient to distort the lattice.

\begin{figure}[tbp]
\includegraphics[width=0.75\columnwidth]{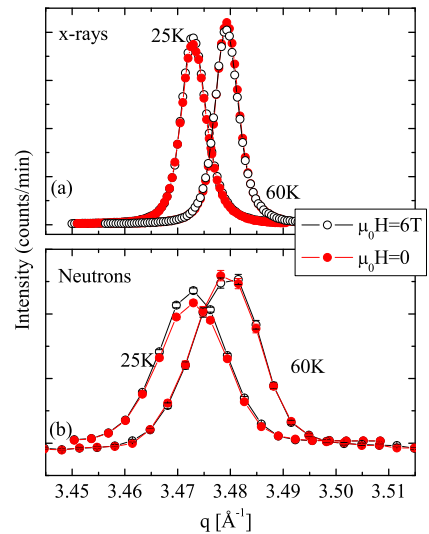}
\caption{(Color online) Field and temperature dependence of the (440) Bragg
peak from (a) X-ray and (b) neutron scattering. The error bars representing
statistical errors of $\pm 1\protect\sigma $ are smaller than the symbols. The peaks in (a) and (b) are resolution limited.}
\label{fig:xrays}
\end{figure}

We also searched for field effects in high resolution X-ray scattering. The
X-ray powder diffraction experiments were conducted at the APS Argonne
National Laboratory on the 11-ID-C beamline. A high-energy, $115$~keV x-ray
beam with a high-resolution analyzer was used with a $6$~T magnet. In order
to dismiss any grain orientation with the field, GE varnish was applied. In
Fig.~\ref{fig:xrays}(a) we plot the most intense cubic $(440)$ peak with and
without an applied field at $60$~K and $25$~K. The peaks are
field-independent in both shape and intensity. Needless to say, no peak
splitting or new peaks were found when the field was applied and moreover the peaks are resolution limited.
This rules out any global structural transition due to temperature or field.

Finally, similar experiments were performed using neutrons which are more
sensitive to scattering from oxygen, but have lower resolution. These
experiments were done on the BT1 powder diffractometer at NIST,
Gaithersburg, USA, with a field up to 6~T perpendicular to the scattering
plane. Data were collected at the same temperatures, with a neutron energy
of $E=34.5$~meV. In Fig. \ref{fig:xrays}(b) we plot the $(440)$ Bragg peak.
As with the x-ray picture, no apparent difference is revealed between the
measurements with and without the field. These peaks are resolution limited as well.
However, the small difference in intensity at $T=25$~K between the $6$~T and zero field measurements is real
and reproducible. It is probably due to very small structure-factor or
magnetic form-factor changes induced by the field.

To conclude, both local probes unambiguously indicate a lattice deformation
takes place as $T$ approaches $T_{f}$. In addition, the magnetization
dependence of the muon spin relaxation rate shows that lattice deformation
is affected by magnetic field, therefore pointing to magneto-elastic
coupling. Long range scattering measurements fail to detect global changes
in the lattice parameters upon application of the field. Since the
distortion is found locally by resonances, but not globally by scattering,
it must be within the unit cell. It affects, at most, the structure-factors
on the one hand, and the hyperfine coupling between Y nuclei and Mo spin on
the other. Our work indicates that magneto-elastic coupling is part of the
freezing process of YMO and provides a simple way to detect it in other
magnets.

We are grateful to the staff of TRIUMF for assistance with the $\mu ^{+}$SR
experiments. OO and AK acknowledge the financial support of NATO -
Collaborative Linkage Grant, reference number PST.CLG.978705, the Israel-US
Binational Science Foundation and the HFM network of the ESF, and the
European Commission under the 6th Framework Programme through the Key
Action: Strengthening the European Research Area, Research Infrastructures.
Contract n$^{\circ }$: RII3-CT-2004-506008. Use of APS was supported by the
U.S. DOE under Contract No. DE-AC02-06CH11357.

\end{document}